# Contact morphology and revisited photocurrent dynamics in monolayer MoS$_2$

Eric Parzinger[1,2], Martin Hetzl[1,2], Ursula Wurstbauer[1,2] and Alexander W. Holleitner[1,2,*]

[1] *Walter Schottky Institute and Physics Department, Technical University of Munich, Am Coulombwall 4a, 85748 Garching, Germany.*

[2] *Nanosystems Initiative Munich (NIM), Schellingstr. 4, 80799 Munich, Germany.*

**Two-dimensional (2D) layered transition metal dichalcogenides (TMDs) have emerged as promising materials for electronic, optoelectronic, and valleytronic applications. Recent work suggests drastic changes of the band gap and exciton binding energies of photo-excited TMDs with ultrafast non-radiative relaxation processes effectively heating the crystal lattice. Such phenomena have not been considered in the context of optoelectronic devices yet. We resolve corresponding ultrafast photoconductance dynamics within monolayer MoS$_2$ and demonstrate that a bolometric contribution dominates the overall photoconductance. We further reveal that a focused laser illumination, as is used in many standard optoelectronic measurements of MoS$_2$, modifies and anneals the morphology of metal contacts. We show that a junction evolves with lateral built-in electric fields, although Raman- and photoluminescence spectra indicate no significant changes such as a crystal phase transition. We highlight how such optimized devices can drive ultrafast electromagnetic signals in on-chip high-frequency and THz circuits.**

**Introduction**

Future applications of transition metal dichalcogenides (TMDs) rely on the fabrication of good and well-defined contacts.[1–24] Although a lot of progress has been made in the fabrication of metal contacts with reduced contact resistance,[3,6,7,10,25,26] the corresponding impact on optoelectronic phenomena in TMDs is not yet fully understood.[4,6] In fact, low resistant contacts remain still the key bottleneck for the realization of a high device performance, and they are especially interesting for optoelectronic devices to avoid depletion regions at the contacts. In general, a combination of top and edge contact is used to describe the morphology of contacts to TMDs.[6] The top contact is formed either by a van der Waals gap between metal and the TMD-crystal or by the creation of covalent bonds to the two-dimensional semiconductor, leading to Fermi level pinning[7,25] or the metallization[6,25] of the TMD in the contact region. As for pure edge contacts, the fabrication turns out to be rather difficult for obvious reasons and so far, has only been reported for graphene.[27] Thus, understanding the interfacial properties of the metal-TMD interface, dominating for most contact geometries, is a key aspect. Considerable efforts have been made by investigating different contact metals by means of density functional theory (DFT) calculations[25,26] as well as the realization of field effect transitors (FETs) with different metals.[3,4,7,10] These studies show, that for monolayer $MoS_2$, Ti/Au contacts exhibit the least contact resistance and a Schottky barrier height of ~200 meV.[3,7,25] However, the influence of the metal contact on the underlying $MoS_2$ remains unclear. This holds in particular, if one considers the presence of defects such as sulfur vacancies at the basal plane of the 2D material. Defects may not only cause Fermi level shifts of up to 1 eV over tens of nanometers,[4] they also open the possibility for the formation of covalent bonds between the contact metal and the TMDs.[6,25] Whereas covalent bonds may reduce the interface resistance, a significant impact on the Fermi level in the 2D material is expected. Recent reports show by evaluating the transfer curves of field effect transistors, that the work function of the contact metal has very weak influence on the Fermi level pinning in $MoS_2$.[7] Noteworthy, the best contacts so far have been achieved by inducing a metallic phase transition within the 2D semiconductors, e.g. via a phase transition from 2H to 1T in $MoS_2$ FETs.[5]

The Coulomb screening in monolayer TMDs can be significantly altered by the presence of a large density of photogenerated charge carriers.[15,28–32] In turn, it has been demonstrated that both the quasi-particle band gap and the excitonic binding energies are renormalized after a pulsed photo-excitation of the TMDs. The renormalization effects are based upon an interplay of excitation-induced dephasing, phase-space filling and the mentioned screening of Coulomb interaction in combination with ultrafast non-radiative relaxation and recombination processes.[15,28–32] Particularly, the latter give rise to a transfer of the photo-induced excess energy to the phonon bath.[32,33] In turn, the lattice heats up to several tens of Kelvin, and the corresponding cooling process can last for hundreds of picoseconds depending on the interfacial thermal conductance to the substrate and to the metal contacts of the TMDs.[32,34] Generally, there have been several reports on photocurrent and photoconductance phenomena in TMDs.[2,8,10,12,13,16,17,21] They mainly discuss the observed phenomena in terms of photo-thermoelectric and photovoltaic effects. However, the above-mentioned, drastic photo-induced ultrafast dynamics of the energy scales and local temperatures have not been considered in the discussion of photoconductance phenomena in TMDs yet.

## Results and Discussion

Our study is based on several tens of different monolayer $MoS_2$-flakes, which are micromechanically exfoliated from bulk crystals and then transferred by an all-dry viscoelastic stamping technique (cf. methods).[35] We utilize either $Si/SiO_2$ or sapphire-substrates for the different spectroscopy methods (cf. methods). Fig. 1a depicts an optical image of a $MoS_2$-flake with varying layer thickness (monolayer - 1L, trilayer - 3L, up to bulk) which is placed on top of a 10-nm-thick Ti contact. To demonstrate the laser annealing effect (cf. Fig. 1b and methods), we focus a laser with varying dwell times onto the flake (cf. dot pattern and black lettering in Fig. 1c). Then, we perform atomic force microscopy (AFM) and Kelvin probe force microscopy (KPFM) to characterize the morphology and the work function change within the same flakes (cf. Figs. 1d, 1e, and methods). The impact of the laser-annealing on the work function of monolayer $MoS_2$ is found to be as large as $\Delta E \sim$ -0.16 eV (lower panel of Fig. 1b, and black lettering in Fig. 1e). A comparative study on thermally annealed monolayers suggests values even up to -0.4 eV (Figs. 2a to 2d, and methods). By utilizing Raman and photoluminescence (PL) measurements during the laser-annealing process, we can exclude metallization or the creation of a phase transition from the 2H to the 1T or 1T' phase[5] for monolayer $MoS_2$ in contact with Ti or Au as possible explanations (cf. Supplemental Figs. S1 and methods). However, we can observe an energy difference of the $A_{1g}$ Raman mode of contacted and non-contacted monolayer $MoS_2$, indicating a change of the charge carrier density[36] in the order of $\Delta n \approx$ -2.8·$10^{12}$ cm$^{-2}$ (cf. Supplemental Fig. S2), which is consistent with a shift of the Fermi energy of $\Delta E \approx$ -0.16 eV. Noteworthy, we find that the work function adjustment by the laser-annealing is a persistent effect. It can be observed by performing the KPFM measurements in ambient conditions one week after the actual laser-annealing. This clearly distinguishes the effect from volatile doping effects, such as the previously observed photogating[36] or work function modulation[37] of monolayer MoS$_2$, which are reversible processes related to adsorbed species and ambient gases. Moreover, we do not observe the work function shift for MoS$_2$-flakes positioned directly on Si/SiO$_2$ or sapphire as substrates, i.e. without metals involved (data not shown). In our understanding, the laser-annealing reduces the van der Waals gap between the contacting metal and the MoS$_2$ (cf. Fig. 1b), resulting in an apparent work function adjustment. We note, however, that this gap alteration must be smaller than about ~1 Å, because it cannot be resolved in our AFM measurements (cf. Fig. 1d).

Fig. 3a summarizes the measured work function of MoS$_2$ before and after the annealing steps, as it is determined relative to the value of Ti in air.[38,39] By the impact of a focused laser, we find that the work function of MoS$_2$ can have a local gradient of ~0.16 eV on ~0.5 µm (cf. lower panel in Fig. 1b). This suggests that after an annealing step, horizontal Ti/MoS$_2$ – MoS$_2$ junctions do not necessarily consist of a Schottky barrier.[7,9,17,18,25,40] Instead, they rather comprise an effective junction with built-in electric fields due to the different doping levels (Figs. 3b, 3c, and 3d). In this respect, 1L MoS$_2$ covered by the contact metal yields a shifted and pinned Fermi-level position, while MoS$_2$ away from the contact is intrinsically *n*-doped,[1,4,8] forming an effective junction with built-in electric fields close to the contact's edge.

We point out that all of our investigated optoelectronic metal/MoS$_2$-monolayer/metal devices (about 20 optoelectronic devices in the course of three years) finally turned into having a contact morphology as sketched in Fig. 3. We explain this observation by the long exposure times to a laser as is necessary for optoelectronic experiments. In other words, scanning photocurrent experiments with a focused laser spot anneal the contacts of metal/MoS$_2$-monolayer/metal devices, and this annealing gives rise to a permanent renormalization of the contact energy landscape.

To resolve the impact of the annealed contacts on the optoelectronic dynamics, we present time-integrated data as well as time-resolved photocurrent data on the same 1L MoS$_2$ (cf. Figs. 4, 5, and S3). We utilize photon energies $E_{photon}$ exceeding the ones of the so-called A- and B-exciton in the MoS$_2$ with absorbance energies of $E_{A-exciton}$ = 1.89 eV and $E_{B-exciton}$ = 2.03 eV in our devices.[41] With an absorbance of ~0.03 at $E_{photon}$ = 2.25 eV (for the time-resolved data in Fig. 5) and a maximum laser fluence of 210 µJcm$^{-2}$, we can estimate the initial maximum photogenerated electron-hole pair density to be in the order of ~1.75 · 10$^{13}$ cm$^{-2}$. This charge carrier density is below the so-called Mott-threshold of an exciton-dissociation,[15] but in a regime where a renormalization of both the quasi-particle band gap and the exciton binding energy occurs. The renormalizations happen on timescales as fast as femtoseconds and prevail up to several hundreds of picoseconds after the photo-excitation.[15,28–32]

Generally, we observe that the time-resolved photocurrent signal $I_{sampling}$ as in Fig. 5 can be separated into four components: $I_1, I_2, I_3$, and the *offset* with different characteristic timescales. The initial photocurrent response $I_1$ (blue in Fig. 5) has a temporal Gaussian shape with a FWHM of ~3 ps, which exceeds the experimental resolution of ~1 ps of the utilized THz-time-domain circuit.[42,43] One possible explanation of this first contribution is the so-called photovoltaic effect, as it is termed in standard time-averaged photocurrent experiments.[2,8,10,12,13,16,17,21] On ultrafast timescales as in Fig. 5, this photovoltaic effect translates to both, the so-called transient displacement current $j_D$ and the lifetime-limited current $j_{lifetime}$. The former is typically written as[44]

$$\boldsymbol{j_D} = \varepsilon_r \varepsilon_0 \, \partial \boldsymbol{E}/dt, \tag{1}$$

with $\varepsilon_r$ ($\varepsilon_0$) the relative (vacuum) permittivity and $\partial \boldsymbol{E}/dt$ the ultrafast change of the local electric field $\boldsymbol{E}$, when the laser pulse impinges on the sample. In principle, this change includes the photogeneration of charge carriers as well as the renormalization of the quasi-particle band gap and the exciton binding energy. Importantly, $j_D$ allows to map-out the polarity of electric fields present in the investigated samples. For a positive (negative) bias voltage $U_{sd}$, we observe that the sign of $I_1$ is negative (positive) in the center of the sample (cf. top panel of Fig. 5e). Since $I_1$ has an opposite sign at the two contacts, this ultrafast contribution corroborates the existence of built-in electric fields at the annealed metal contacts. The 'temporal duration' of $j_D$ is given by the dielectric response $\varepsilon_r$ of the overall circuit, which is fundamentally limited by a phonon-excitation in the underlying sapphire substrate at ~12 THz;[43] i.e., significantly faster than the observed constant FWHM of ~3 ps across the whole device. This consideration brings the lifetime-limited current $j_{lifetime}$ into play[44]

$$\boldsymbol{j}_{\text{lifetime}} = \sigma_{\text{eff}} \cdot e^{-\tau/\tau_1} \cdot \boldsymbol{E}, \qquad (2)$$

with $\sigma_{eff}$ the effective conductivity of the locally photo-excited electron bath and $\tau_1$ a corresponding lifetime. From equation (2), we find the same sign for the built-in electric fields at the laser-annealed contacts as in the case of $j_D$, as it is consistent with the sketch in Fig. 3d. The notation of $\sigma_{eff}$ implies that the locally excited electron bath is thermalized, as it can be assumed to occur at ~100s of femtoseconds.[32] At the utilized laser fluences of ~100 µJcm$^{-2}$ and beyond, it has been demonstrated that the non-radiative relaxation and recombination lifetimes are only a few picoseconds.[32] In turn, there is not a clear timescale separation between $j_D$ and $j_{lifetime}$, and we interpret the observed timescale of ~3ps with the non-radiative relaxation and recombination times[32,45] and fast capture mechanisms by traps via Auger processes[46] or surface defects.[47] Consistent with this interpretation, we observe that the FWHM of $I_1$ increases with a lower laser fluence (data not shown).[32] Furthermore, we can neglect a photo-thermoelectric current for this first contribution $I_1$, as we do not detect it for a laser excitation below the band gap of MoS$_2$ (cf. supplementary Fig. S4). In other words, within the given experimental noise, $I_1$ seems to comprise all ultrafast non-radiative processes involving the non-equilibrium dynamics and energetics of charge carriers, which are photogenerated above band gap.

The second contribution $I_2$ shows a fast rise followed by an exponential decay $\propto e^{-\Delta t/\tau_2}$ with $\tau_2 \sim$ 100 ps (cf. orange contribution in Fig. 5a), pointing towards a lifetime-limited current as in equation (2). Since $I_2$ is most pronounced at the contact regions (second panel in Fig. 5e), it also points towards the presence of electric fields as discussed for $I_1$. In this respect, we associate $I_2$ to photoexcited carriers having an extended lifetime $\tau_2$. This interpretation agrees well with the timescale of an energy transfer between the electron bath and the phonon bath,[32] but also with carriers captured by defects via Auger processes and slow trap states[47] or photo-injected charge carriers in indirect-gap side valleys.[48] We note, however, that $I_2$ can be detected for a laser excitation below band gap as well (cf. supplementary Fig. S4). Then, the signal is ~10-fold reduced but with an equivalent timescale. In this respect, $I_2$ comprises a thermo-electric contribution, which is consistent with the fact that the heated electron bath can drive corresponding thermoelectric currents at the contacts' interfaces.[13]

The third component $I_3$ (green in Fig. 5), differs from the previous two contributions by the clearly visible rise time $\tau_3^{\text{rise}}$ followed by a very slow decay time $\tau_3^{\text{decay}}$. Moreover, $I_3$ can be observed both at the contact region (◄) as well as in the center (♦) of the monolayer MoS$_2$ flakes. We attribute $I_3$ to a bolometric effect, which is driven by a heated phonon bath. As demonstrated in Supplemental Fig. S5, the annealed samples exhibit a positive bolometric coefficient $\alpha_{\text{annealed}}$ even for small temperature changes. Taking $\alpha_{\text{annealed}}$ (cf. Supplemental Fig. S5) and the amplitude of $I_{photo}$ (cf. Fig. 4), we can estimate a maximum temperature increase of the overall device to be $\Delta T$ ~60 K. To estimate the maximum temperature increase $\Delta T$ of just the crystal lattice, we consider the specific heat capacitance of monolayer MoS$_2$ $c_{MoS_2}$ and compute $\Delta T = \alpha F / c_{MoS_2}$, where $\alpha$ is the absorption coefficient. For our experimental conditions, $F = 210$ µJ/cm$^2$, $E_{photon} = 2.25$ eV, $\alpha = 0.03$ [49] and $c_{MoS_2} = 4.25 \cdot 10^{-8}$ J cm$^{-2}$ K$^{-1}$,[50] we calculate $\Delta T$ ~150 K. The in-plane thermal conductance of monolayer MoS$_2$ is rather poor and can therefore be neglected.[50,51] Thus, we can estimate the cooling time by $\tau_c = c_{MoS_2}/G$, where $G$ is the interfacial Kapitza conductance. However, the value for $G$ at a van der Waals interface varies strongly between ~0.1 and several 10s of MW m$^{-2}$ K$^{-1}$,[32] which results in timescales ranging from ~10 ps to several nanoseconds. In this interpretation, $\tau_3^{decay}$ corresponds to the time it takes for the monolayer MoS$_2$ on top of sapphire to dissipate the heat introduced by the laser pulse. In this respect, the *offset*-contribution is the prevailing bolometric photoconductance with a cooling time in the order of the laser repetition time (~13.2 ns = 76$^{-1}$ MHz$^{-1}$). We note that a heated phonon bath contributes two-fold to the bolometric photoconductance in TMDs; i.e by a combined change of the mobility and charge carrier density as is known for semiconductors (cf. Supplementary Fig. S5), and by a temperature-induced change of the quasi-particle band gap.[33] Consistently, Fig. 5 demonstrates that $I_3$ and the *offset* are the dominating optoelectronic contributions long after the initial laser pulse, and that their lateral dependences largely agree with the one of the time-integrated photocurrent signal (cf. two lower panels of Figs. 5e and Fig. 4e).

The combined observations let us conclude that scanning photocurrent experiments with a focused laser spot anneal the contacts of metal/MoS$_2$-monolayer/metal devices, and this annealing gives rise to more transparent contacts which favor a bolometric response to dominate the optoelectronic signal. The built-in electric fields at the annealed contacts additionally give rise to photovoltaic currents ($I_1$ and majority of $I_2$), which enhance the overall optoelectronic response close to the contacts. Our results are relevant to a large class of experiments on MoS$_2$ and 2D materials, since a focused laser spot is standard in state-of-the-art optoelectronic characterization techniques.[2,8–14,16–18,20–22,40] Moreover, our work demonstrates how such devices can be implemented into on-chip high-frequency and THz circuits.

## Methods

**Sample Fabrication.** For KPFM, Raman-, and photoluminescence-measurements, the samples are fabricated on a 525-µm-thick *p*-type Si substrate with 285-nm SiO$_2$ layer. On top, Ti contacts with a thickness of 10 nm or Ti/Au contacts with thickness of 5 nm/15 nm are defined by standard photolithography. In a final step, the MoS$_2$-flakes are placed on top of the metal layer. This inverted geometry allows us to directly access the physical properties of the MoS$_2$-flakes by the mentioned KPFM, Raman- or photoluminescence-methodologies. The samples for temperature-dependent current-voltage (IV) traces are based on a 525-µm-thick sapphire substrate. Here, monolayers of MoS$_2$ are first transferred and subsequently contacted via Ti/Au contacts with a thickness of 5 nm/20 nm. For the photocurrent characterization, we utilize a 430-µm-thick sapphire substrate covered with a 300-nm-thin layer of ion-implanted Si. In a first optical lithography step, an Auston-switch[44] geometry is formed via HNO$_3$/HF etching (cf. Supplementary Fig. S3). The remaining silicon strip serves as an on-chip photodetector (Auston-switch) as we use for ultrafast photocurrent measurements.[42,43,52] In a second optical lithography step, we evaporate 10 nm/300 nm Ti/Au to define the contacts in form of two coplanar striplines. The striplines contact the monolayer MoS$_2$ and the read-out of the Auston-switch. Each individual stripline has a width of 5 µm and a distance of 10 µm in-between two coplanar ones. The MoS$_2$-flakes are placed at a distance of about 400 µm from the Auston-switch. The striplines have a total length exceeding 58 mm. The geometry allows for both time-integrated and time-resolved photocurrent measurements on individual MoS$_2$-flakes.

**Optical microscopy.** All samples are characterized by optical microscopy to investigate the layer-thickness and morphology of the individual MoS$_2$-flakes. Fig. 1a depicts such an optical image of a MoS$_2$-flake with varying layer thickness placed on top of a 10-nm-thick Ti contact (on a 525-µm-thick *p*-type Si substrate with 285-nm SiO$_2$ layer). The monolayer, trilayer, and bulk areas have different contrast and they are marked as 1L, 3L and bulk, respectively.

**Laser-annealing.** We perform laser-annealing by continuously illuminating individual spots on the samples to locally introduce heat to the MoS$_2$-flakes on Ti (Fig. 1b). For the presented results, we use a Kr/Ar-laser with a wavelength of 488 nm, which is focused with a 100x objective onto the sample stored at ~10$^{-3}$ mbar. However, we note that we achieve similar results with a Ti:Sapphire laser at 780 nm. In both cases, the laser spot is scanned across the samples by the help of a x-y-z-piezo scanner (PI) with a specific dose per laser spot position. Fig. 1c introduces a corresponding dose pattern as is applied to the MoS$_2$-flake shown in Fig. 1a. At each of the depicted positions (gray dots in Fig. 1c), the laser is focused at a dose of 0.4 kJ/cm$^2$ forming a uniform grid with 1 µm spacing. Moreover, a dose gradient is superimposed to the pattern in the shape of a 'T-U-M' logo (shaded dark area in Fig. 1c). There, the dose ranges from ~10$^4$ to ~5×10$^4$ J/cm$^2$. We adjust the laser dose by varying the dwell time of each spot illuminated with a power of ~1.5 mW. Fig. 2d illustrates the measured work function shift induced by laser-annealing as a function of the overall applied laser doses. The doses are again adjusted by varying the dwell time with a laser wavelength of 488 nm, a laser power of 1.5 mW, and a spot diameter of ~1 µm. The corresponding KPFM traces are extracted across a laser-annealed spot and fitted by a single Gaussian (cf. Fig. 1b). In Fig. 2d, the amplitude of the Gaussian is plotted as a measure for the maximum shift. We find that the work function shift seems to saturate at ~ -0.2 eV (red line in Fig. 2d).

**Atomic force microscopy.** All atomic force microscopy (AFM) measurements are performed in ambient conditions using a Bruker MultiMode 8 atomic force microscope by means of PeakForce frequency-modulation with a SiN cantilever and a Si tip with a triangular shape.[53] Fig. 1d depicts the AFM image of the MoS$_2$-flake on top of the Ti contact after laser-annealing. In the topography profile of the sample, there are no indications from the laser-annealing step. Only small protruding clusters are visible on the MoS$_2$-flake, which can be attributed to residual contaminations introduced by the viscoelastic stamping technique.[35]

**Kelvin probe force microscopy.** Kelvin probe force microscopy (KPFM) measurements are carried out in ambient conditions using the same system as for AFM measurements in dual pass mode (lift height 50 nm).[53] Fig. 1e shows a KPFM map carried out simultaneously to the AFM results in Fig. 1d. Clearly, the 'T-U-M' pattern is visible in the KPFM signal of the MoS$_2$-flake placed on the thin Ti film. We detect that the work function is shifted by up to $\Delta E \approx -(0.1 - 0.2)$ eV by the laser-annealing for the highest laser dose (Fig. 1e and lower panel in Fig. 1b). Noteworthy, we observe that the monolayer regions, which are not laser-annealed, show no difference in work function. This observation indicates that prior to laser-annealing, the van der Waals gap between the MoS$_2$-flake and the Ti contact is rather large, causing no significant Fermi level pinning.[6,7] Given the fact that Ti is often used to promote the adhesion of Au contacts, the question arises if Ti or Au induces an additional work function shift of 1L MoS$_2$. Therefore, we directly compare the Fermi level pinning of monolayer MoS$_2$ on Ti and Au before and after a thermal annealing step, as described in the next section.

**Thermal annealing.** Fig. 2a depicts an AFM image of monolayer MoS$_2$ (encompassed by a red solid line) transferred on top of a 10-nm-thick Ti- (left) and a 15-nm-thick Au-contact (right) with a central overlap region (~1 µm) of Ti on top of Au. For the Au-contact, a 5-nm-thick Ti is used below the Au, to promote the adhesion of the Au. In a second step, KPFM-traces of the monolayer MoS$_2$ and of the bare contact metals are extracted of the as-fabricated samples. The traces are taken along the dashed (1) and dotted (2) lines in Fig. 2a and depicted in Figs. 2b and 2c, respectively. For initialization of the KPFM-traces, the work function of Ti in air is set to be -4.33 eV.[38] We are aware of possible changes due to oxidation of the Ti while exposed to ambient conditions, however, this can cause both negative and positive changes.[39] As we are focusing on the relative work function shifts, we neglect any offset of the absolute values caused by oxide formation. For MoS$_2$ on Ti, we find then roughly -4.45 eV and on Au -4.25 eV, suggesting a change of the work function difference due to the different metals by ~0.2 eV. In average, these values agree well with the previously reported work function of monolayer MoS$_2$ on top of *p*-type Si by KPFM measurements also in ambient conditions[37]. In a third step, the sample is thermally annealed for 45 min at 200 °C and in a vacuum of ~1 mbar. Figs. 2b and 2c show the corresponding KPFM traces after the thermal annealing step with the individual shift of the work functions indicated by red arrows. We find a significant difference in work function of the MoS$_2$-flake before and after the annealing (red arrows Fig. 2b). Within the overlap region, the 10-nm-thick Ti on top of the Au contact seems to define the work function of MoS$_2$. Fig. 2c shows the KPFM traces of the bare metals (along the dotted line (2) in Fig. 2a). For Ti, the work function increases slightly, while for Au the work function slightly decreases (indicated by arrows). We interpret the small changes to stem from the removal of lithography residues during the thermal annealing.

**Raman and photoluminescence spectroscopy.** Raman and photoluminescence spectroscopy are powerful tools to probe a variety of fundamental properties of MoS$_2$ such as the number of layers,[41,54] charge carrier density,[36] charge carrier lifetime,[45,47,48] temperature,[51] etc. We utilize Raman and photoluminescence spectroscopy to clearly identify the number of layers for the region of interest for all presented samples, and most importantly, we can exclude thermal degradation or an annealing induced phase transition (cf. Supplemental Fig. S1).[5]

**Temperature-dependent transport spectroscopy.** We take current-voltage (IV)-characteristics on annealed two-terminal metal-MoS$_2$-metal devices over a temperature range from 305 to 325 K (cf. Supplementary Fig. S5). In the devices, the Ti/Au-contacts (5/20 nm) cover the monolayer MoS$_2$-flake from the top on a sapphire-substrate (cf. Supplemental Figs. S2 and S5). All IV-measurements are performed using an Oxford Instruments Optistat setup to control the temperature at a pressure of ~$10^{-5}$ mbar. The source-drain current is recorded utilizing a current voltage amplifier at a gain of $10^7$ (cf. Supplemental Fig. S5).

**Scanning photocurrent spectroscopy.** Scanning photocurrent measurements are performed utilizing a Ti-Sapphire laser together with a non-linear optical fiber to photoexcite the MoS$_2$-flakes.[2,8,10,12,13,16,17,21] The photon energy is set to be $E_{photon}$ = 2.25 eV at a pump fluence $F$ = 210 µJ/cm$^2$, a repetition rate 76 MHz, and a pulse duration ~1 ps. The samples are kept at room temperature and vacuum conditions at a pressure of $10^{-5}$ mbar throughout the optoelectronic measurements. Before the presented photocurrent measurements, all samples are laser-annealed. We record the photocurrent signal by chopping the laser beam at a frequency of 1.7 kHz together with a lock-in detection. A current voltage amplifier is utilized to amplify the signal before lock-in detection. Fig. 4a depicts an optical microscopy image of a monolayer MoS$_2$-flake contacted by two metal striplines. For better visibility, the MoS$_2$-flake is highlighted by solid white lines. For detecting photocurrent maps, the laser spot with diameter ~2.5 µm is scanned across the sample while the photocurrent response as well as the reflectivity are simultaneously recorded for each spot. Fig. 4b shows such a reflectivity map of the MoS$_2$-flake, while Figs. 4c and 4d show the corresponding photocurrent maps for $U_{sd}$ = -5 V and +5 V. In the figures, the solid blue lines indicate the position of the MoS$_2$-flake, while the dashed gray lines highlight the positions of the contacting metal striplines. In the electronic circuitry, the right stripline represents the source contact, while the left stripline is the drain contact connected to ground. For negative (positive) bias, we observe a pronounced negative (positive) photocurrent signal close to the right (left) metal stripline (cf. Figs. 4c and 4d). For better visualization, Fig. 4e shows a cross-section of the photocurrent signal $I_{photo}$ for both $U_{sd}$ = -5 V and +5 V (along the dashed-dotted line illustrated in Figs. 4b-d). The cross-section of the reflectivity is also shown with a maximum reflectivity at the striplines (gray lines). We clearly observe that the time-integrated photocurrent is mainly generated at the contact regions but there are also finite smaller contributions in the center region. Noteworthy, for zero bias $U_{sd}$ = 0 V, we observe a photocurrent response at both contacts with equal amplitude but opposite sign (data not shown). All time-averaged data are consistent with recent reports on the photoconductance and photocurrent phenomena of TMDs.[2,8,10,12,13,16,17,21]

**Ultrafast photocurrent spectroscopy.** For measuring the ultrafast optoelectronic dynamics of the monolayer MoS$_2$-flakes, we use an on-chip THz-time domain photocurrent detection scheme based on coplanar striplines[34,42,43,52,55,56] (cf. Supplemental Fig. S3). A short laser pulse with $E_{photon}$ = 2.25 eV at a pump fluence $F$ = 210 µJ/cm$^2$, a repetition rate 76 MHz, and a pulse duration of < 1ps is used to excite charge carriers in the MoS$_2$. The corresponding photocurrent response in the MoS$_2$-flake gives rise to electromagnetic transients in the striplines, which propagate along the striplines. We utilize a time-delayed laser pulse in combination with an Auston-switch[44] as an on-chip read-out of the electromagnetic transients. The probe laser pulse has a photon energy of 1.59 eV and a temporal width of 100 fs. The current $I_{sampling}$ across the Auston-switch samples the electromagnetic transients on the striplines as a function of the time-delay $\Delta t$ between the two laser pulses. It is directly proportional to the ultrafast photocurrents in $MoS_2$. Hereby, the photocurrents in $MoS_2$ can be measured with a (sub-) picosecond time-resolution[43]. Fig. 5 depicts the time-resolved photocurrent $I_{sampling}(\Delta t)$ of the monolayer $MoS_2$ excited at the drain contact (Figs. a and b) and at the center of the channel (Figs. c and d) on short (0-250 ps) and long (0-1600 ps) timescales. We can consistently fit the time-resolved data $I_{sampling}$ with four components: an initial ultrafast response $I_1$ with a Gaussian lineshape and a FWHM ~ 3 ps (blue in Fig. 5), a second response $I_2$ as a Gaussian convoluted decay-function with a decay time $\tau_2$ (orange in Fig. 5), a third comparably long lasting response $I_3$ with a rise time $\tau_3^{rise}$ and a decay time $\tau_3^{decay}$ (green in Fig. 5) and an *offset*. The *offset* is particularly observable for $\Delta t < 0$; shortly before the pump laser hits the $MoS_2$-flake in the repetitive scheme with a repetition time of 76 $MHz^{-1}$ ~13.2 ns. Fig. 5e depicts the integrated area of the fitting functions for all four contributions as a function of position along the $MoS_2$-flake for $U_{sd} = +5$ V (open symbols) and for $U_{sd} = -5$ V (filled symbols).

**References**

1. Radisavljevic, B., Radenovic, A., Brivio, J., Giacometti, V. & Kis, A. Single-layer $MoS_2$ transistors. *Nat. Nanotechnol.* **6,** 147–150 (2011).

2. Wang, Q. H., Kalantar-Zadeh, K., Kis, A., Coleman, J. N. & Strano, M. S. Electronics and optoelectronics of two-dimensional transition metal dichalcogenides. *Nat. Nanotechnol.* **7,** 699–712 (2012).

3. Das, S., Chen, H.-Y., Penumatcha, A. V. & Appenzeller, J. High Performance Multilayer $MoS_2$ Transistors with Scandium Contacts. *Nano Lett.* **13,** 100–105 (2013).

4. McDonnell, S., Addou, R., Buie, C., Wallace, R. M. & Hinkle, C. L. Defect-Dominated Doping and Contact Resistance in $MoS_2$. *ACS Nano* **8,** 2880–2888 (2014).

5. Kappera, R. *et al.* Phase-engineered low-resistance contacts for ultrathin $MoS_2$ transistors. *Nat. Mater.* **13,** 1128–1134 (2014).

6. Allain, A., Kang, J., Banerjee, K. & Kis, A. Electrical contacts to two-dimensional semiconductors. *Nat. Mater.* **14,** 1195–1205 (2015).

7. Kim, C. *et al.* Fermi Level Pinning at Electrical Metal Contacts of Monolayer Molybdenum Dichalcogenides. *ACS Nano* **11,** 1588–1596 (2017).

8. Yin, Z. *et al.* Single-Layer $MoS_2$ Phototransistors. *ACS Nano* **6,** 74–80 (2012).

9. Buscema, M. *et al.* Large and Tunable Photothermoelectric Effect in Single-Layer $MoS_2$. *Nano Lett.* **13,** 358–363 (2013).

10. Fontana, M. *et al.* Electron-hole transport and photovoltaic effect in gated $MoS_2$ Schottky junctions. *Sci. Rep.* **3,** 1634 (2013).

11. Sundaram, R. S. *et al.* Electroluminescence in Single Layer $MoS_2$. *Nano Lett.* **13,** 1416–1421 (2013).

12. Lopez-Sanchez, O., Lembke, D., Kayci, M., Radenovic, A. & Kis, A. Ultrasensitive photodetectors based on monolayer $MoS_2$. *Nat. Nanotechnol.* **8,** 497–501 (2013).

13. Baugher, B. W. H., Churchill, H. O. H., Yang, Y. & Jarillo-Herrero, P. Optoelectronic devices based on electrically tunable p-n diodes in a monolayer dichalcogenide. *Nat. Nanotechnol.* **9,** 262–267 (2014).

14. Ross, J. S. *et al.* Electrically tunable excitonic light-emitting diodes based on monolayer $WSe_2$ p-n junctions. *Nat. Nanotechnol.* **9,** 268–272 (2014).

15. Steinhoff, A., Rösner, M., Jahnke, F., Wehling, T. O. & Gies, C. Influence of Excited Carriers on the Optical and Electronic Properties of $MoS_2$. *Nano Lett.* **14,** 3743–3748 (2014).

16. Furchi, M. M., Polyushkin, D. K., Pospischil, A. & Mueller, T. Mechanisms of Photoconductivity in Atomically Thin $MoS_2$. *Nano Lett.* **14,** 6165–6170 (2014).

17. Zhang, Y. *et al.* Photothermoelectric and photovoltaic effects both present in $MoS_2$. *Sci. Rep.* **5,** (2015).

18. Hong, T. *et al.* Plasmonic Hot Electron Induced Photocurrent Response at $MoS_2$–Metal Junctions. *ACS Nano* **9,** 5357–5363 (2015).

19. Koperski, M. *et al.* Single photon emitters in exfoliated $WSe_2$ structures. *Nat. Nanotechnol.* **10,** 503–506 (2015).

20. Li, H. *et al.* Optoelectronic crystal of artificial atoms in strain-textured molybdenum disulphide. *Nat. Commun.* **6,** 7381 (2015).

21. Wang, H., Zhang, C., Chan, W., Tiwari, S. & Rana, F. Ultrafast response of monolayer molybdenum disulfide photodetectors. *Nat. Commun.* **6,** 8831 (2015).

22. Mak, K. F. & Shan, J. Photonics and optoelectronics of 2D semiconductor transition metal dichalcogenides. *Nat. Photonics* **10,** 216–226 (2016).

23. Mak, K. F., He, K., Shan, J. & Heinz, T. F. Control of valley polarization in monolayer $MoS_2$ by optical helicity. *Nat. Nanotechnol.* **7,** 494–498 (2012).

24. Zeng, H., Dai, J., Yao, W., Xiao, D. & Cui, X. Valley polarization in $MoS_2$ monolayers by optical pumping. *Nat. Nanotechnol.* **7,** 490–493 (2012).

25. Kang, J., Liu, W., Sarkar, D., Jena, D. & Banerjee, K. Computational Study of Metal Contacts to Monolayer Transition-Metal Dichalcogenide Semiconductors. *Phys. Rev. X* **4,** 031005 (2014).

26. Popov, I., Seifert, G. & Tománek, D. Designing Electrical Contacts to $MoS_2$ Monolayers: A Computational Study. *Phys. Rev. Lett.* **108,** 156802 (2012).

27. Wang, L. *et al.* One-Dimensional Electrical Contact to a Two-Dimensional Material. *Science* **342,** 614–617 (2013).

28. Chernikov, A., Ruppert, C., Hill, H. M., Rigosi, A. F. & Heinz, T. F. Population inversion and giant bandgap renormalization in atomically thin $WS_2$ layers. *Nat. Photonics* **9,** 466–470 (2015).

29. Schmidt, R. *et al.* Ultrafast Coulomb-Induced Intervalley Coupling in Atomically Thin $WS_2$. *Nano Lett.* **16,** 2945–2950 (2016).

30. Ulstrup, S. *et al.* Ultrafast Band Structure Control of a Two-Dimensional Heterostructure. *ACS Nano* **10,** 6315–6322 (2016).

31. Pogna, E. A. A. *et al.* Photo-Induced Bandgap Renormalization Governs the Ultrafast Response of Single-Layer $MoS_2$. *ACS Nano* **10,** 1182–1188 (2016).

32. Ruppert, C., Chernikov, A., Hill, H. M., Rigosi, A. F. & Heinz, T. F. The Role of Electronic and Phononic Excitation in the Optical Response of Monolayer $WS_2$ after Ultrafast Excitation. *Nano Lett.* **17,** 644–651 (2017).

33. Waldecker, L. *et al.* Momentum-Resolved View of Electron-Phonon Coupling in Multilayer $WSe_2$. *ArXiv170303496 Cond-Mat* (2017).

34. Brenneis, A. *et al.* Ultrafast electronic readout of diamond nitrogen-vacancy centres coupled to graphene. *Nat. Nanotechnol.* **10,** 135–139 (2015).

35. Castellanos-Gomez, A. *et al.* Deterministic transfer of two-dimensional materials by all-dry viscoelastic stamping. *2D Mater.* **1,** 011002 (2014).

36. Miller, B., Parzinger, E., Vernickel, A., Holleitner, A. W. & Wurstbauer, U. Photogating of mono- and few-layer MoS$_2$. *Appl. Phys. Lett.* **106,** 122103 (2015).

37. Lee, S. Y. *et al.* Large Work Function Modulation of Monolayer MoS$_2$ by Ambient Gases. *ACS Nano* **10,** 6100–6107 (2016).

38. *CRC handbook of chemistry and physics: a ready-reference book of chemical and physical data*. (CRC, 1998).

39. Ohler, B., Prada, S., Pacchioni, G. & Langel, W. DFT Simulations of Titanium Oxide Films on Titanium Metal. *J. Phys. Chem. C* **117,** 358–367 (2013).

40. Yamaguchi, H. *et al.* Spatially Resolved Photoexcited Charge-Carrier Dynamics in Phase-Engineered Monolayer MoS$_2$. *ACS Nano* **9,** 840–849 (2015).

41. Wurstbauer, U., Miller, B., Parzinger, E. & Holleitner, A. W. Light–matter interaction in transition metal dichalcogenides and their heterostructures. *J. Phys. Appl. Phys.* **50,** 173001 (2017).

42. Brenneis, A. *et al.* THz-circuits driven by photo-thermoelectric, gate-tunable graphene-junctions. *Sci. Rep.* **6,** 35654 (2016).

43. Kastl, C., Karnetzky, C., Brenneis, A., Langrieger, F. & Holleitner, A. Topological Insulators as Ultrafast Auston Switches in On-Chip THz-Circuits. *IEEE J. Sel. Top. Quantum Electron.* **23,** 1–5 (2017).

44. Auston, D. H., Johnson, A. M., Smith, P. R. & Bean, J. C. Picosecond optoelectronic detection, sampling, and correlation measurements in amorphous semiconductors. *Appl. Phys. Lett.* **37,** 371–373 (1980).

45. Korn, T., Heydrich, S., Hirmer, M., Schmutzler, J. & Schüller, C. Low-temperature photocarrier dynamics in monolayer MoS$_2$. *Appl. Phys. Lett.* **99,** 102109 (2011).

46. Shi, H. *et al.* Exciton Dynamics in Suspended Monolayer and Few-Layer MoS$_2$ 2D Crystals. *ACS Nano* **7,** 1072–1080 (2013).

47. Wang, H., Zhang, C. & Rana, F. Ultrafast Dynamics of Defect-Assisted Electron–Hole Recombination in Monolayer MoS$_2$. *Nano Lett.* **15,** 339–345 (2015).

48. Docherty, C. J. *et al.* Ultrafast Transient Terahertz Conductivity of Monolayer MoS$_2$ and WSe$_2$ Grown by Chemical Vapor Deposition. *ACS Nano* **8,** 11147–11153 (2014).

49. Funke, S. *et al.* Imaging spectroscopic ellipsometry of MoS$_2$. *J. Phys. Condens. Matter* **28,** 385301 (2016).

50. Saha, D. & Mahapatra, S. Analytical insight into the lattice thermal conductivity and heat capacity of monolayer MoS$_2$. *Phys. E Low-Dimens. Syst. Nanostructures* **83,** 455–460 (2016).

51. Yan, R. *et al.* Thermal Conductivity of Monolayer Molybdenum Disulfide Obtained from Temperature-Dependent Raman Spectroscopy. *ACS Nano* **8,** 986–993 (2014).

52. Seifert, P., Vaklinova, K., Kern, K., Burghard, M. & Holleitner, A. Surface State-Dominated Photoconduction and THz Generation in Topological Bi$_2$Te$_2$Se Nanowires. *Nano Lett.* **17,** 973–979 (2017).

53. Hetzl, M., Kraut, M., Hoffmann, T. & Stutzmann, M. Polarity Control of Heteroepitaxial GaN Nanowires on Diamond. *Nano Lett.* **17,** 3582–3590 (2017).

54. Mak, K. F., Lee, C., Hone, J., Shan, J. & Heinz, T. F. Atomically Thin MoS$_2$: A New Direct-Gap Semiconductor. *Phys. Rev. Lett.* **105,** 136805 (2010).

55. Prechtel, L. *et al.* Time-resolved ultrafast photocurrents and terahertz generation in freely suspended graphene. *Nat. Commun.* **3,** 646 (2012).

56. Kastl, C., Karnetzky, C., Karl, H. & Holleitner, A. W. Ultrafast helicity control of surface currents in topological insulators with near-unity fidelity. *Nat. Commun.* **6,** (2015).

**Acknowledgements**

This work was supported by a European Research Council (ERC) Grant NanoREAL (No. 306754). We further acknowledge financial support by Deutsche Forschungsgemeinschaft (DFG) via excellence cluster 'Nanosystems Initiative Munich' (NIM), through the TUM International Graduate School of Science and Engineering (IGSSE) and BaCaTeC.

**Author Contributions**

The manuscript was written through contributions of all authors. All authors have given approval to the final version of the manuscript. E.P., U.W., and A.W.H. designed the experiments, E.P., and M.H. performed the experiments.

**Conflict of Interest**

The authors declare no competing financial interest.

**Supporting Information Available**

Supplementary Information accompanies this paper at www.nature.com/npj2dmaterials. Reprints and permission information is available online at http://npg.nature.com/reprintsandpermissions/. Correspondence and requests for materials should be addressed to A.W.H.

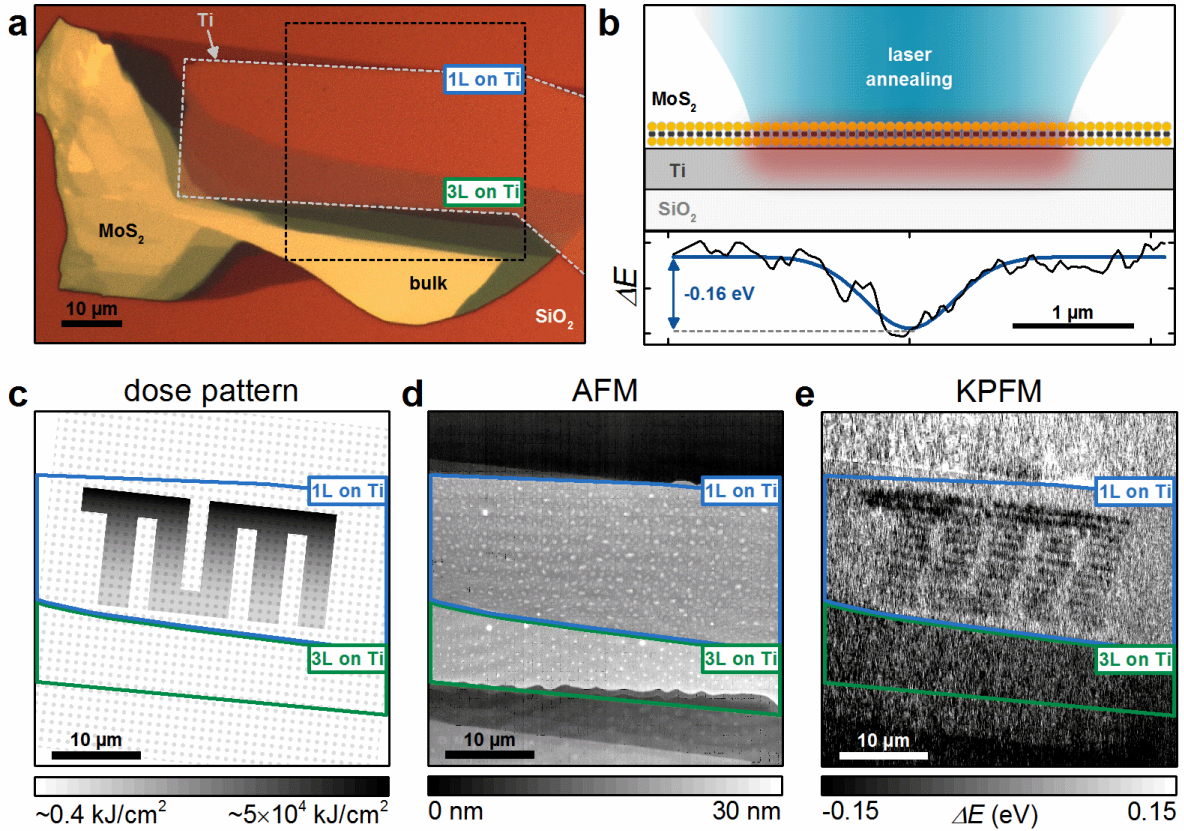

**Fig. 1. Laser processing of MoS₂-on-titanium contacts. a**, Optical microscopy image (100x) of an exfoliated mono- (1L), tri- (3L) and bulk MoS₂-flake stamped on top of a 10-nm-thick titanium (Ti) contact. **b**, Schematic side view of the Ti-contact with the MoS₂-flake on top with a focused laser beam. The laser-induced change of the work function $\Delta E$ of a monolayer MoS₂ on top of 10-nm shows a Gaussian shape with maximum of $\Delta E = -0.16$ eV. The total laser dose is 0.74 GJ/cm² at $E_{photon} = 2.54$ eV. **c**, Similar laser dose pattern applied to the black dashed area in Fig. 1a at power of ~1.5 mW. Each dot in (c) represents a laser spot position where the laser energy is applied for a certain dwell time amounting to 0.4 kJ/cm² per dot. Moreover, the shaded area highlights an increased laser dose ranging from ~$10^4$ to ~$5 \times 10^4$ J/cm². **d**, Atomic force microscope (AFM) image of the dashed area in Fig. 1a after laser patterning. **e**, Corresponding Kelvin probe force microscope (KPFM) image after laser patterning. The lettering pattern with an increased laser dose can be clearly recognized as an area with a lowered work function. In Figs. **c, d** and **e,** the solid blue (green) lines mark the 1L (3L) part of the MoS₂-flake on top of the Ti-contact.

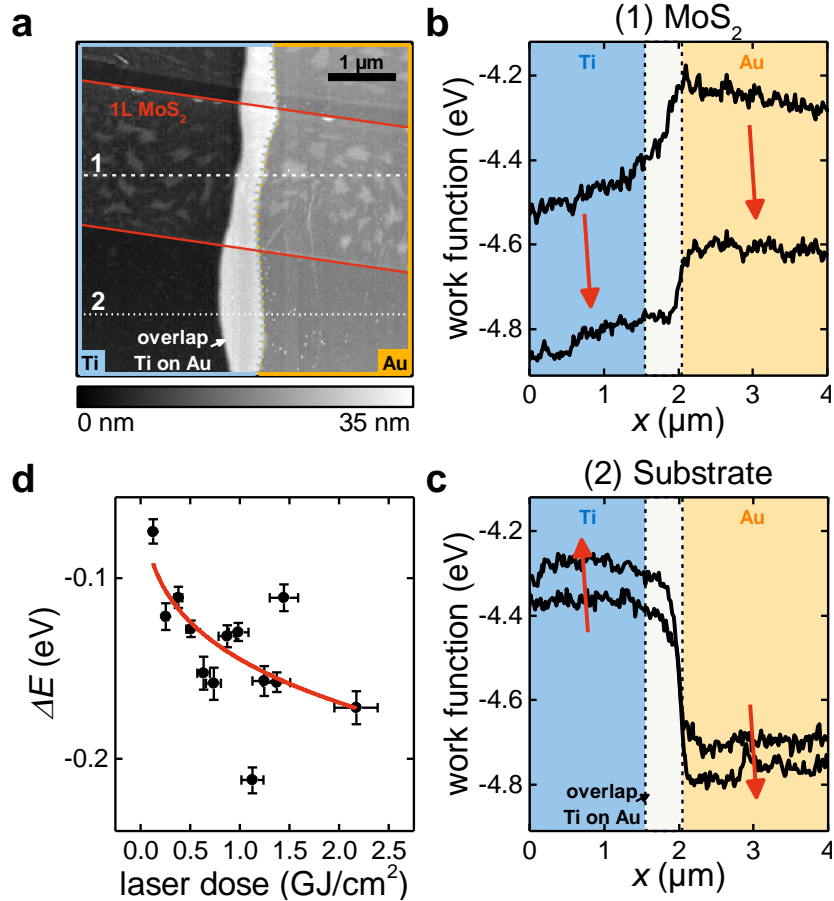

**Fig. 2. Work function mapping for temperature-treated MoS₂-metal contacts and comparison to laser processing. a,** AFM image of a monolayer MoS$_2$-flake (marked by dotted red lines) on top of a 10-nm-thick Ti-contact (left) and a 15-nm-thick Au-contact (right). The adhesion of Au is promoted by 5 nm Ti. In the central overlap region (bright), Ti is on top of Au. The image area is $5 \times 5$ µm$^2$. **b,** Work function profile of the MoS$_2$-flake along the dashed line (1) in Fig. 2a with a clear decrease of the work function after thermal annealing at 200 °C for 45 min in a vacuum of 1 mbar. Arrows indicate the downshift of the respective work function due to annealing. **c,** Equivalent work function profile for the Ti- and Au-contacts along the dotted line (2) in Fig. 2a with a upshift (downshift) for Ti (Au) after annealing. For initialization, the work function of Ti in air is defined to be -4.33 eV (cf. methods). **d,** Laser-induced change of work function $\Delta E$ of a monolayer MoS$_2$ on top of 10-nm Ti-contact.

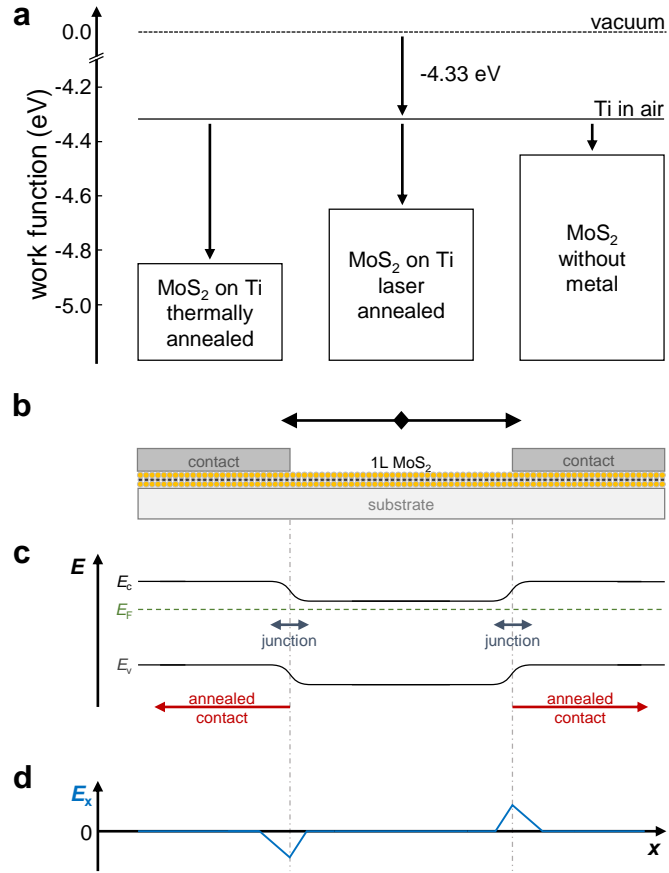

**Fig. 3. Sketch of the effective band structure of a laser processed MoS$_2$-metal contact. a,** Experimetally determined work functions of monolayer MoS$_2$ in the pristine case and after a thermal or a laser annealing step with respect to the value of Ti in air. **b,** The contact comprises Au and Ti on top of MoS$_2$-monolayers on an sapphire-substrate (alternatively SiO$_2$/Si-substrates). **c,** Band structure sketched across the contact region with $E_C$ ($E_V$) the energy of the conduction (valence) band of the MoS$_2$-flake and $E_F$ the Fermi energy. The laser processing downshifts $E_F$ at the position of the metal contact. As a result, an effective junction with adjacent areas of differing Fermi levels evolves. **d,** Sketch of the resulting built-in electric field $E_x$ at the interface between the laser-annealed contact region and pristine MoS$_2$.

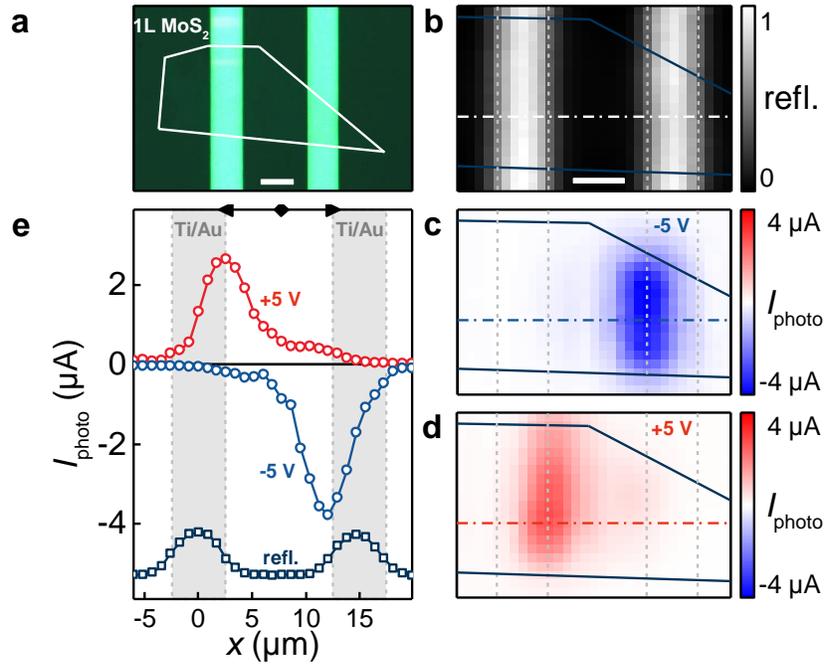

**Fig. 4. Time-averaged optoelectronic response in MoS₂-devices with laser processed contacts.**
**a**, Optical microscope image (100x) of a monolayer MoS$_2$-flake (encompassed by solid lines) and two Ti/Au-contacts (10/300 nm) forming striplines on a sapphire substrate. **b**, Reflectivity map of the investigated central area of the device with solid lines highlighting the area of the MoS$_2$-flake. Scale bars, 5 µm. **c**, Concurrently measured map of the time-averaged photocurrent $I_{photo}$ at $E_{photon}$ = 2.25 eV, when the laser is swept across the same area as in Fig. 4b for a bias voltage of $U_{sd}$ = -5 V. **d,** Equivalent photocurrent map for $U_{sd}$ = +5 V. **e**, Time-integrated photocurrent $I_{photo}$ for $U_{sd}$ = +5 V and -5 V vs the *x*-direction (perpendicular to the striplines) along the dotted-dashed lines in Figs. 4b-d. On the bottom of the figure, the reflectivity signal is shown (not to scale). Gray areas mark the position of the striplines.

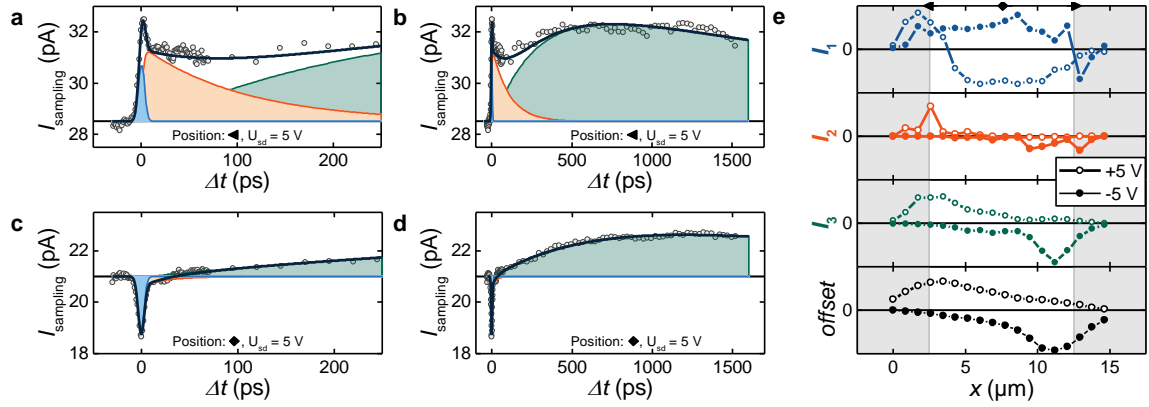

**Fig. 5. Ultrafast optoelectronic response in two-terminal MoS$_2$-devices with laser processed contacts. a**, Time-resolved photocurrent $I_{\text{sampling}}(\Delta t)$ of the monolayer MoS$_2$-device as in Fig. 4 on a picosecond timescale at $E_{\text{photon}} = 2.25$ eV. The laser excitation occurs at the drain contact (at position $x = 2.6$ µm in Fig. 5e) with a laser spot with FWHM ~2.5 µm and $U_{\text{sd}} = +5$ V. **b,** Same measurement for a timescale up to 1600 ps. (**c,d**) Equivalent measurements at the center of the channel ($x = 7.8$ µm as in Fig. 5e). The fits to the three distinct contributions $I_1$ (blue), $I_2$ (orange) and $I_3$ (green) are highlighted in color. **e**, The integrated area of the three contributions $I_1$, $I_2$, $I_3$ and the initial *offset* to the photocurrent signal vs the position for a bias voltage of + 5 V (open circles) and - 5 V (filled circles). Gray areas mark the position of metal contacts.

# Contact morphology and revisited photocurrent dynamics in monolayer MoS$_2$

Eric Parzinger[1,2], Martin Hetzl[1,2], Ursula Wurstbauer[1,2] and Alexander W. Holleitner[1,2]

[1] *Walter Schottky Institute and Physics Department, Technical University of Munich, Am Coulombwall 4a, 85748 Garching, Germany.*

[2] *Nanosystems Initiative Munich (NIM), Schellingstr. 4, 80799 Munich, Germany.*

- Supporting material -

**Supplementary Notes**

**1. Supplemental Note on transport spectroscopy.** Supplemental Fig. S5 depicts temperature dependent IV-traces of the device depicted in Supplemental Fig. S2 on a logarithmical scale. In particular, Supplemental Fig. S5a shows IV-traces after laser-annealing of the metal contacts, while Supplemental Fig. S5b prior. The change of the current due to temperature is color coded and indicated by arrows. Supplemental Figs. S5c and S5d illustrate the temperature dependence of the source-drain current exemplarily shown for a drain-source bias $U_{sd} = 5$ V. The current $I$ vs temperature is fitted assuming a linear temperature dependence (red lines in Supplemental Figs. S5c and S5d)

$$I(T) = I(T_0)\,(1 + \alpha \Delta T), \tag{1}$$

with $\alpha$ a linear bolometric temperature coefficient and $\Delta T = T - T_0$ the temperature difference with respect to $T_0 = 305$ K. The analysis yields $\alpha_{\text{annealed}} = (+0.011 \pm 0.001)\,\text{K}^{-1}$ after the laser-annealing and $\alpha_{\text{prior}} = (-0.018 \pm 0.002)\,\text{K}^{-1}$ prior to the annealing. We note that after annealing (Supplemental Fig. S5a), the rather symmetric IV characteristics can be understood in terms of a metal-semiconductor-metal device, in which the semiconducting temperature dependence of the monolayer $MoS_2$ dominates with a positive bolometric coefficient $\alpha_{\text{annealed}}$. In contrast, the as-fabricated device (i.e. without annealing) shows rather asymmetric IV traces together with a factor ~10 higher overall device resistance (Supplemental Fig. S5b). We attribute both findings to a poor contact interface between the monolayer $MoS_2$ and at least one of the metal leads, resulting in irregular charge injection properties. Microscopically, this can be related to residuals at the interface, resulting in a larger van der Waals gap as discussed before[6]. Moreover, we find a negative bolometric coefficient $\alpha_{\text{prior}}$ prior to annealing (Supplemental Fig. S5d), which we can explain, if we consider that the overall device characteristic is governed by the metal leads and the poor contacts, rather than by the monolayer $MoS_2$.

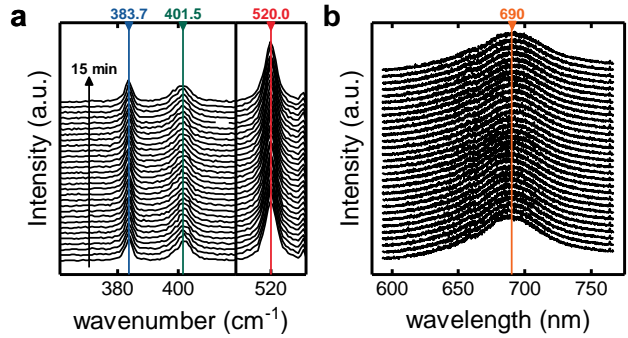

**Supplemental Fig. S1. Raman and photoluminescence (PL) of MoS$_2$ on Ti. a,b**, Raman and PL signal of monolayer MoS$_2$ recorded during laser-annealing. ($E_{photon}$ = 2.54 eV, 1.5 mW, ~10$^{-3}$ mbar, 30 s integration time each).

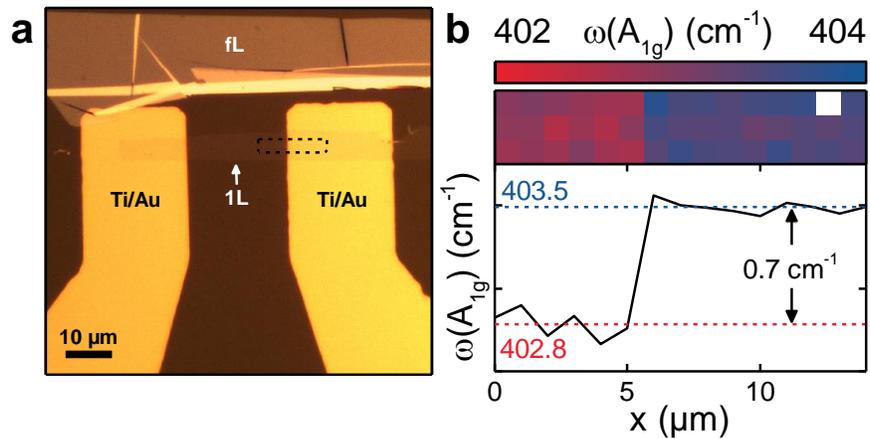

**Supplemental Fig. S2. Raman signal of monolayer MoS$_2$ beside a Ti/Au metal contact and underneath the contact. a**, Optical microscope image of the two-terminal device of a monolayer MoS$_2$-flake. In the device, the Ti/Au-contacts (5/20 nm) cover the monolayer MoS$_2$-flake from the top on an sapphire-substrate. **b**, Energy map (top) and binned trace (bottom) of the A$_{1g}$ Raman mode taken from the area marked by the dashed rectangle in Fig. a. The trace illustrates an energy shift of the A$_{1g}$ mode by $\Delta\omega(A_{1g}) \approx 0.7$ cm$^{-1}$ between uncontacted (left) and contacted (right) monolayer MoS$_2$.

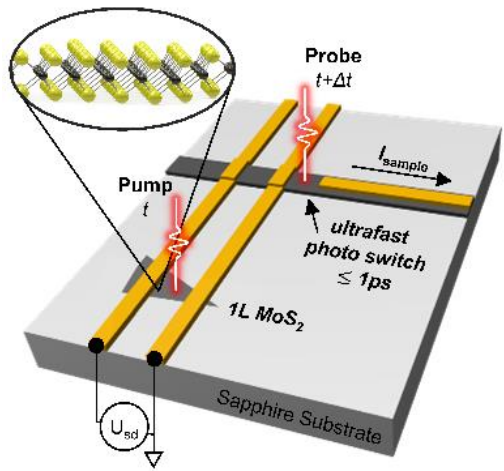

**Supplemental Fig. S3. Ultrafast photocurrent read-out scheme.** Device geometry and optoelectronic pump-probe circuit. The pump laser pulse is focused on the monolayer $MoS_2$-flake contacted by the stripline circuit. The probe pulse triggers the sampling circuit. The Ti/Au-electrodes are depicted in yellow, while the ion-implanted silicon (Si) is represented in dark gray. The latter forms the ultrafast photo switch (Auston-switch) to measure the time-resolved photocurrent $I_{sampling}$ as a function of the time delay $\Delta t$.[31,43,65,66]

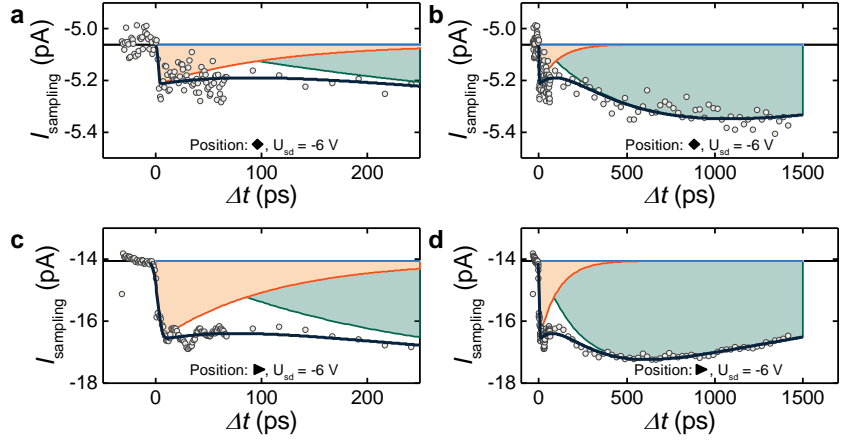

**Supplemental Fig. S4. Below band-gap excitation: ultrafast photocurrents in two-terminal MoS₂-devices with laser processed contacts. a**, Time-resolved photocurrent $I_{\text{sampling}}(\Delta t)$ of the monolayer MoS$_2$-device as in Fig. 5 on a picosecond timescale at $E_{\text{photon}} = 1.68$ eV. The laser excitation occurs at the center of the channel $x = 8.0$ µm in Fig. 5e) with a laser spot with FWHM ~2.5 µm and $U_{\text{sd}} = -6$ V. **b,** Same measurement for a timescale up to 1600 ps. (**c,d**) Equivalent measurements at the source contact ($x = 11.6$ µm as in Fig. 5e). The fits to the tree distinct contributions include $I_1$, $I_2$ (orange) and $I_3$ (green) are highlighted in color, although we note that $I_1$ is not resolvable within the given noise; i.e. we cannot detect $I_1$ for an excitation below the band gap of MoS$_2$.

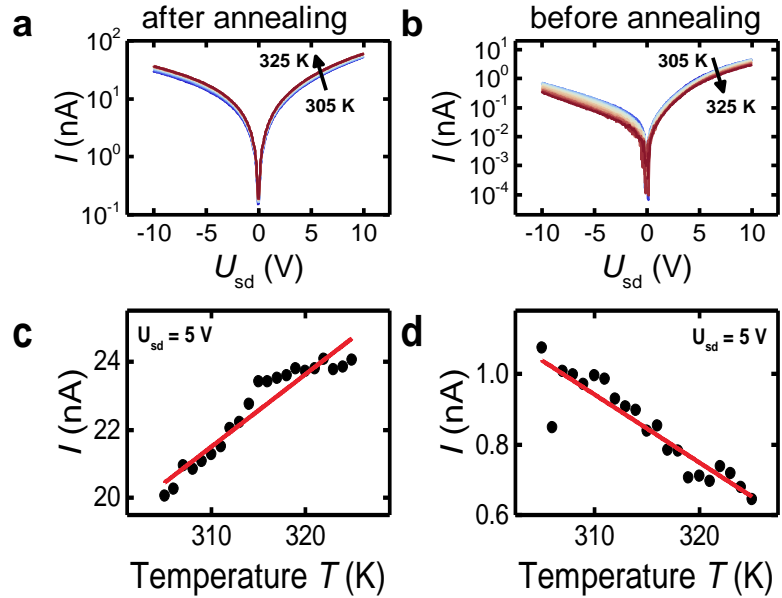

**Supplemental Fig. S5. Electronic characteristics of a MoS$_2$-device with laser-processed contacts. a**, Temperature-dependent current-voltage (IV) traces of a two-terminal metal-monolayer-MoS$_2$-metal device after laser-annealing of the contacts. The device is depicted in Supplementary Figure S2a. **b,** Corresponding IV-traces before laser-annealing of the contacts. (**c,d**) Dark-current $I$ at a bias voltage of $U_{sd} = +5V$ as a function of the bath temperature after and prior to laser-annealing. Lines are fits that highlight the altered bolometric response after and before laser-processing of the contacts.